\documentclass[runningheads]{llncs}

\usepackage{geometry}
\usepackage[utf8]{inputenc} 
\usepackage[T1]{fontenc}    
\usepackage{hyperref}       
\usepackage{url}            
\usepackage{booktabs}       
\usepackage{nicefrac}       
\usepackage{microtype}      
\usepackage{lipsum}
\usepackage{amsmath,amssymb,amsfonts}
\usepackage{algorithm}
\usepackage{cite}
\usepackage{bm}
\usepackage{amsmath}
\usepackage{algorithmic}
\usepackage{graphicx}
\graphicspath{ {Figures/} }
\usepackage{stfloats}
\usepackage{multicol}
\usepackage{multirow}
\usepackage{tablefootnote}
\usepackage[flushleft]{threeparttable}
\usepackage{color, colortbl}
\usepackage{lipsum}
\usepackage{siunitx}
\usepackage{tablefootnote}
\usepackage{mathptmx}

\usepackage{array}
\newcommand{\PreserveBackslash}[1]{\let\temp=\\#1\let\\=\temp}
\newcolumntype{C}[1]{>{\PreserveBackslash\centering}p{#1}}
\newcolumntype{R}[1]{>{\PreserveBackslash\raggedleft}p{#1}}
\newcolumntype{L}[1]{>{\PreserveBackslash\raggedright}p{#1}}

\definecolor{Gray}{gray}{0.9}
\definecolor{Highlight}{rgb}{0.99, 0.99, 0.9}

\begin{document}
\title{A Low-Power Dual-Factor Authentication Unit \\ for Secure Implantable Devices}

\author{
  Saurav Maji\inst{1} \and Utsav Banerjee\inst{1} \and Samuel H. Fuller\inst{1,2} \and Mohamed R. Abdelhamid\inst{1} \and\\ Phillip M. Nadeau\inst{2} \and Rabia Tugce Yazicigil\inst{3} \and Anantha P. Chandrakasan\inst{1}}

\authorrunning{Maji et al.}
%
\institute{
Massachusetts Institute of Technology, Cambridge, MA USA  
\and Analog Devices Inc., Boston, MA USA
\and Boston University, Boston, MA USA}
\maketitle              
\begin{abstract}
This paper presents a dual-factor authentication protocol and its low-power implementation for security of implantable medical devices (IMDs). The protocol incorporates  traditional cryptographic first-factor authentication using Datagram
Transport Layer Security – Pre-Shared Key (DTLS-PSK) followed by the user's touch-based voluntary second-factor authentication for enhanced security. With a low-power compact always-on wake-up timer and touch-based wake-up circuitry, our test chip consumes only 735 pW idle state power at 20.15 Hz and 2.5 V. The hardware accelerated dual-factor authentication unit consumes 8 \bm{$\mu$}W at 660 kHz and 0.87 V. Our test chip was coupled with commercial Bluetooth Low Energy (BLE) transceiver, DC-DC converter, touch sensor and coin cell battery to demonstrate standalone implantable operation and also tested using in-vitro measurement setup.

\end{abstract}

\section{Introduction}

Implantable medical devices (IMDs), such as cardiac defibrillators, pacemakers, cochlear implants, bladder stimulators and pressure sensors, are widely used to improve the quality of lives of patients. Recent advances in microelectronics and medical technology have enabled Internet-connected IMDs which can be controlled by the patients/users through external handheld or wearable devices. However, several proof-of-concept attacks have been demonstrated on such devices by exploiting weaknesses in authentication protocols or their implementations \cite{zhang_imd_2014}. While such connected implantable devices have the potential to enable many emerging medical applications such as on-command implantable drug delivery, security concerns pose a threat to their widespread deployment. To address this challenge, we present a secure low-power IC with sub-nW sleep-state power, energy-efficient cryptographic acceleration and a novel dual factor authentication mechanism which ensures that the ultimate security of the IMD lies in the hands of the user.

\section{Dual-Factor Authentication Protocol}

A typical wireless implantable system (exemplified by on-command drug-delivery system) comprises the following parties -- the user, their IMD and a secure cloud server. The user requires the IMD to perform a desired action as per their command (e.g. deliver drug with specified dosage). To securely execute this action, the server must authenticate the IMD and then communicate the drug dosage and associated parameters as set by the physician. The server performs the following critical functions:
\begin{itemize}
  \item Authenticate the IMD before the execution of any action.
  \item Approve legitimate commands by the user, e.g., the server verifies whether the drug dose requested by the user lies within the prescribed limits set by the physician.
  \item Maintain a record of transactions between the user and the server, which can be accessed by the physician later for diagnostic purposes.
\end{itemize}
However, the resource-constrained IMD cannot directly communicate with the server. Therefore, a cellphone is used as a relay between the IMD and the server. The IMD connects with the cellphone using Bluetooth Low Energy (BLE) and the cellphone is connected to the server through the Internet. To secure the authentication and the communication scenario, we consider the following threat model: 

\begin{enumerate}
    \item An adversary having control over the cellphone through malware, ransomware or physical possession can force the IMD to execute malicious actions without the user's knowledge.  Traditional single factor cryptographic-based authentication protocols are not sufficient to prevent such attacks since the IMD cannot distinguish whether it has received a command from a legitimate user or an adversary.
    \item If the cellphone is also used to wake up the IMD (through the BLE connection), an adversary having control over the cellphone can also attempt to wake up the system repeatedly and drain its energy, requiring frequent battery replacement.   
\end{enumerate}

\begin{figure}[!b]
  \centering
  {\includegraphics[width=12.5 cm]{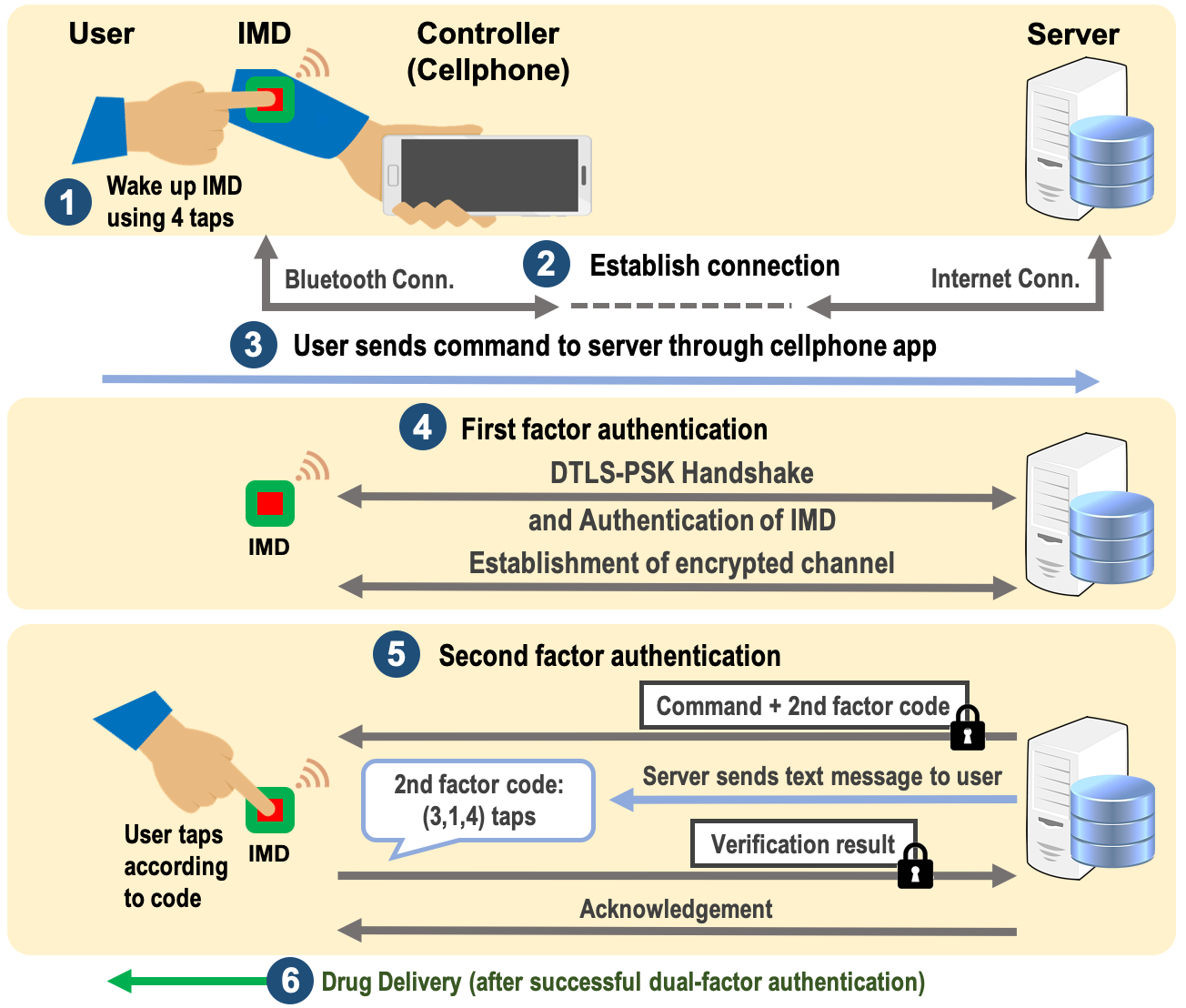}}
  \caption{Proposed dual-factor authentication protocol to secure internet-connected implantable medical devices (IMDs).}
  \label{fig:protocol}
\end{figure}

As a solution, we propose a dual-factor authentication scheme where cryptographic authentication is supplemented with a voluntary response from the user \cite{maji_thesis_2019}. The voluntary response serves as a guarded action from the user, that is, it represents consent from the user for executing the desired action without causing much inconvenience to the user. In our protocol, we have selected a touch-based voluntary response where the user taps on their skin near the IMD. Since most implants are subcutaneous, they can easily detect the tap-pattern and authenticate using this second-factor response. Clearly, it is difficult for an adversary to provide correct second-factor response to the IMD without alerting the user, hence providing higher security guarantees. In addition to second factor authentication, the human voluntary factor (human touch) is also used for waking up the system. This provides dual benefits of achieving extremely low-power wake-up as well as protection against energy-drainage attacks.\\

As shown in Fig. \ref{fig:protocol}, our proposed dual-factor authentication protocol consists of the following steps:
\begin{enumerate}
  \item The user follows a pre-defined tap pattern (e.g. 4 taps) to wake up the IMD.
  \item The IMD, after being woken up, connects with the server through the user’s cellphone acting as a relay.
  \item The user enters their credentials through the cellphone and connects to the server.
  \item The server performs first factor authentication using Datagram Transport Layer Security – Pre-Shared Key (DTLS-PSK) \cite{rescorla_tls1p3_2018, eronen_psk_2005, badra_psk_2009}. The DTLS-PSK protocol provides end-to-end security using only symmetric cryptography and the handshake involves exchange of only $\sim 200$ bytes of data, making it ideal for low-power applications \cite{banerjee_eedtls_2017}. The pre-shared key (PSK), used for authentication, is known only to the IMD and the server, both configured when the IMD was implanted inside the user.
  \item The server sends a text message to the user with a random tap-pattern which acts as a one-time password, and the same pattern is also sent to the IMD through the DTLS-encrypted channel along with control information. The user follows this pattern to perform the touch-based second factor response which gets verified by the IMD. The IMD notifies the server about the authentication result, which the server acknowledges back. The server also informs the user about the authentication result through another text message.
  \item Upon successful authentication of both factors, the IMD executes drug delivery and then goes back to the idle state.
\end{enumerate}
During every handshake, the DTLS-PSK protocol derives fresh encryption keys from the PSK and additional randomness obtained from handshake messages \cite{eronen_psk_2005}. This ensures the use of different keys for each session and thus avoiding any potential side-channel attacks. 

\section{System Architecture}

\begin{figure}[!t]
  \centering
  {\includegraphics[width=13 cm]{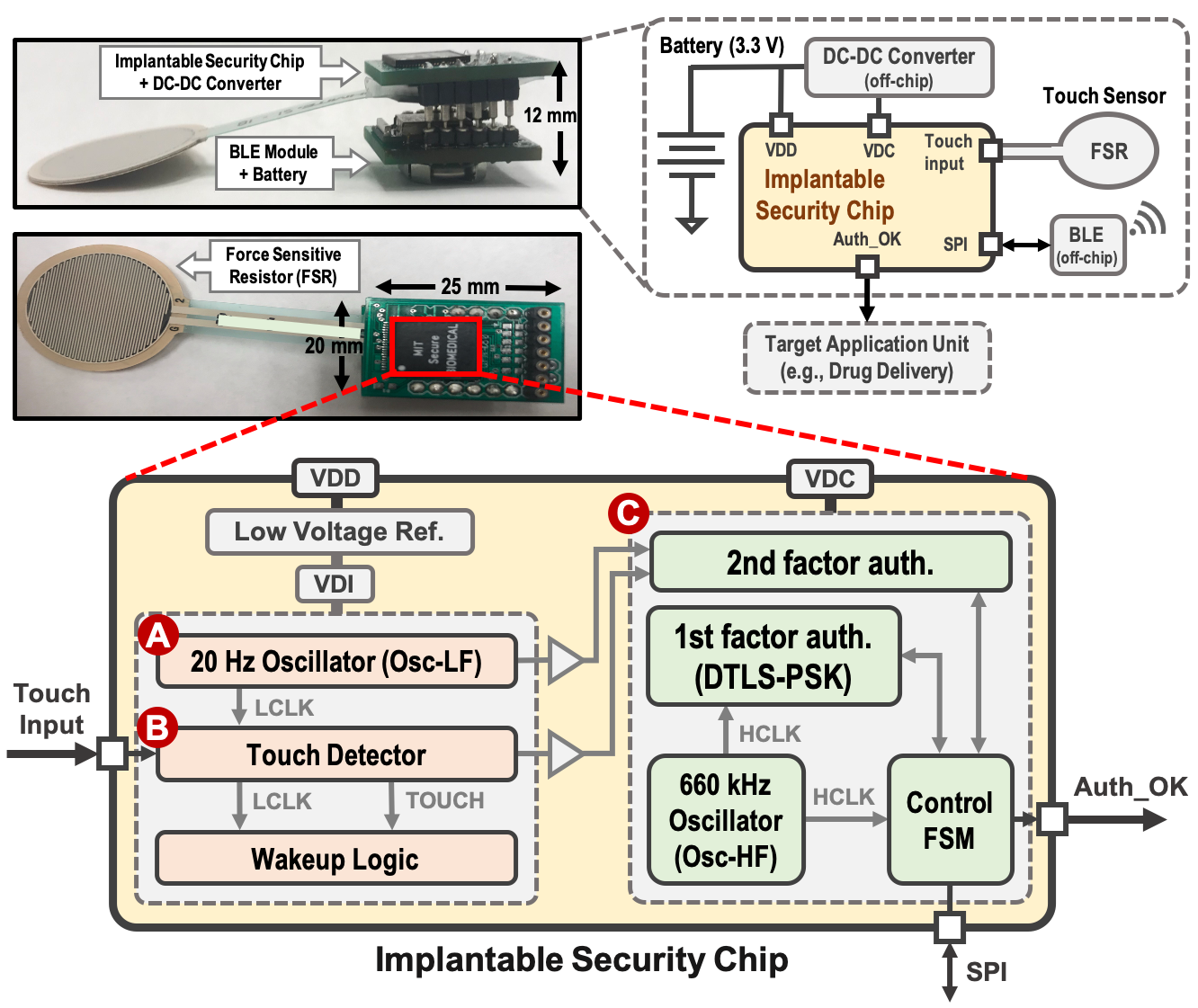}}
  \caption{Prototype secure implantable device with system components and details of security chip architecture.}
  \label{fig:architecture}
\end{figure}

Fig. \ref{fig:architecture} shows our implantable security chip architecture, which consists of three main circuit blocks -– (A) an always-on wake-up timer, (B) a touch detection circuit to recognize tap patterns for wake-up and second factor response and (C) an authentication block which performs post-wake-up dual factor authentication. The complete test system, powered by a battery and a DC-DC converter, consists of an off-chip Bluetooth Low Energy (BLE) module to handle wireless communications and a force sensitive resistor (FSR) to detect tap patterns, connected to our test chip through SPI (Serial Peripheral Interface) and touch input pins respectively. As an example, this test setup can be integrated with an implantable drug-delivery unit to securely authenticate its operation. 

\subsection{Always-On Wake-Up Timer}

\begin{table}[!b]
\renewcommand{\arraystretch}{1.25}
\caption{Comparison of Core Oscillator with State-of-the-Art Hz-Range Oscillators}
\label{table:oscillator}
\centering
\begin{tabular}{|l|C{15mm}|C{15mm}|C{15mm}|C{15mm}|C{15mm}|}
\hline
\rowcolor{Gray}
{\textbf{Metric}}  & {\cite{lin_osc_2009}} & {\cite{nadeau_osc_2016}} & {\cite{wang_osc_2016}}  & \multicolumn{2}{c|}{\textbf{This Work}}  \\
\hline
{Process (nm)} & {180} & {180} & {65} & \multicolumn{2}{c|}{65}   \\
\hline
{Voltage (V)} & {0.6} & {0.6} & {0.5} & {0.4} & {0.85}   \\
\hline
{Freq (CKA) (Hz)} & {11.11} & {18} & {2.8} & {1.36} & {10.07} \\
\hline
{Power (pW)} & {100} & {4.2} & {44.4} & {0.246} & {9.59} \\
\hline
{Energy (pJ/cyc)} & {9.0} & {0.23} & {15.8} & {0.181} & {0.952} \\
\hline
{Area ($\mu$m\textsuperscript{2})} & {19000} & {180000} & {25500} & \multicolumn{2}{c|}{745} \\
\hline
\end{tabular}
\end{table}

\begin{figure}[!t]
  \centering
  {\includegraphics[width=13 cm]{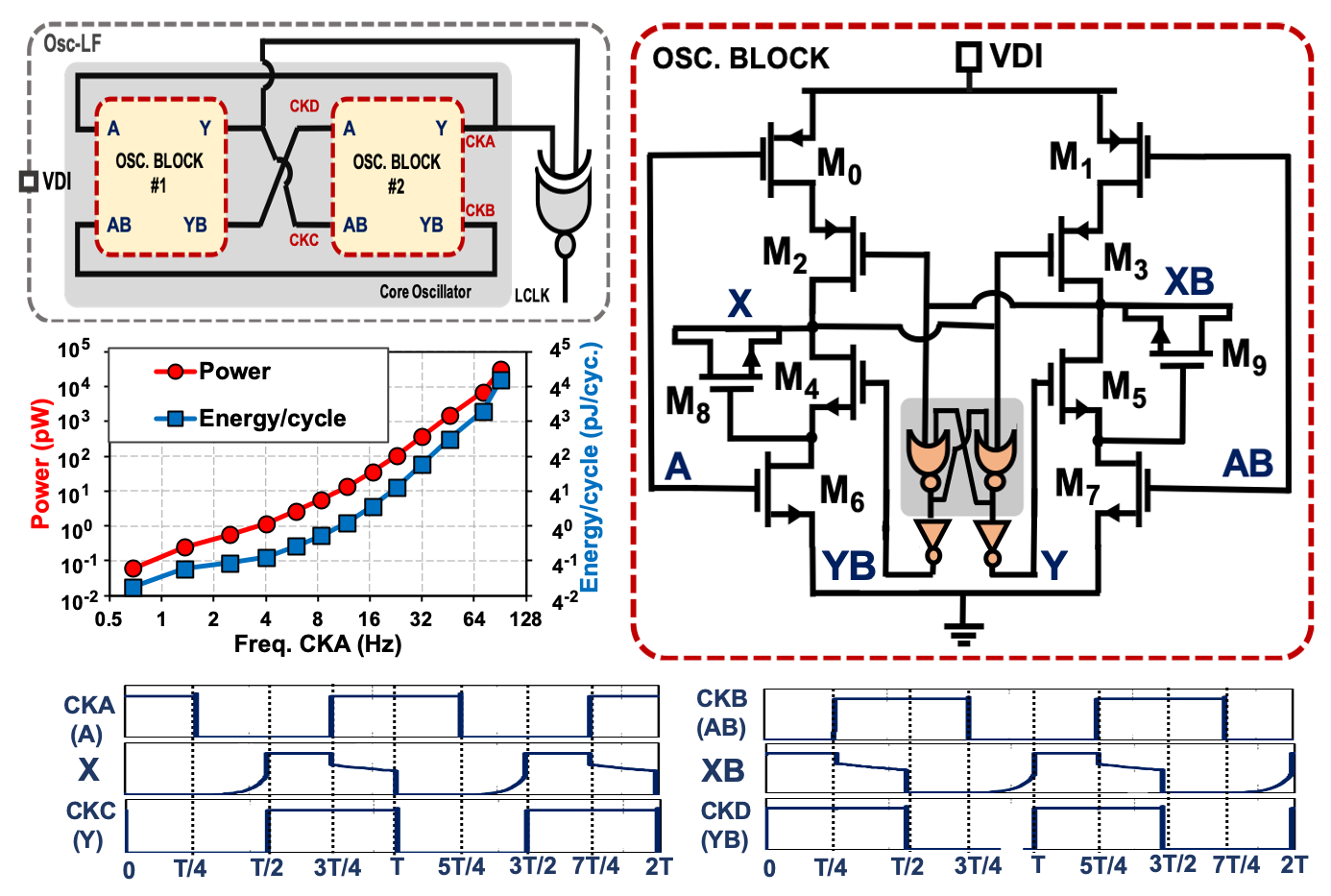}}
  \caption{Topology and characteristics of the low-frequency oscillator (Osc-LF), with simulation waveform of nodes in OSC BLOCK \#1 (marked above). }
  \label{fig:osc}
\end{figure}

The always-on wake-up unit needs to have sub-nW power consumption because of its continuous operation, thus motivating the need for an energy-efficient timer. The timer must also occupy small area for resource-constrained IMD. In our design, a gate-leakage based timer is chosen because of its low power consumption. The wide variation of gate-leakage current with temperature is not a major concern to our design because the temperature inside human body is well regulated. Also, inaccuracies in timing do not affect operation as its output is only used for wake-up purposes. Fig. \ref{fig:osc} shows the topology of our low-frequency oscillator (Osc-LF)  with the following key design optimizations:
\begin{enumerate}
  \item The core oscillator topology (marked gray in Fig. \ref{fig:osc}) comprises two back-to-back identical self-timed stages, each providing self-resetting logic to the other. The reduction in number of stages leads to reduction in area.
  \item Each stage consists of a two-transistor gate-leakage-based differential structure for generating the time constant. This further removes the need for any external current/voltage references leading to further area savings.  
  \item The incorporation of stack effect at every node removes short-circuit current and provides power savings.  
\end{enumerate}

The core oscillator, when characterised separately, runs at 1.36 Hz with 246 fW power consumption at 0.4 V. As shown in Table \ref{table:oscillator}, it achieves better power and power-efficiency with significantly lower area as compared to 
\cite{lin_osc_2009},
\cite{nadeau_osc_2016} and \cite{wang_osc_2016}. The four-phased clock generated by the core oscillator is further used to double the output frequency at the node LCLK using an XOR gate. The supply voltage (VDI) is tuned to 0.85 V to obtain LCLK frequency 20.15 Hz (CKA frequency 10.07 Hz, shown in Table \ref{table:oscillator}) for the IMD operation.  

\subsection{Touch-Detection Unit}

\begin{figure}[!t]
  \centering
  {\includegraphics[width=13 cm]{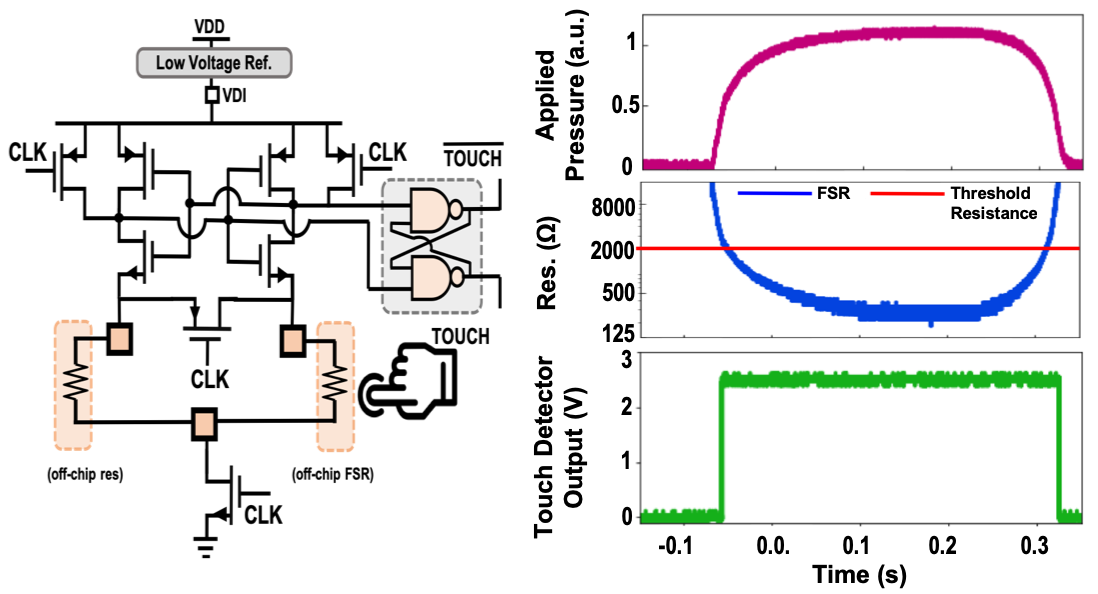}}
  \caption{Topology and characteristics of the touch detection circuit.}
  \label{fig:touch}
\end{figure}

The touch detection circuit, shown in Fig. \ref{fig:touch}, uses a commercially available Force Sensitive Resistor (FSR) which changes its resistance in response to applied pressure. The resistance of this sensor, which will also be implanted under the skin, is compared with an appropriate fixed resistance (2 k\si{\ohm}) to detect the user’s tap patterns. The difference in resistance of the pull-down path causes regenerative feedback to produce a rail-to-rail differential output, thus consuming only dynamic power. 

\begin{figure}[!t]
  \centering
  {\includegraphics[width=13 cm]{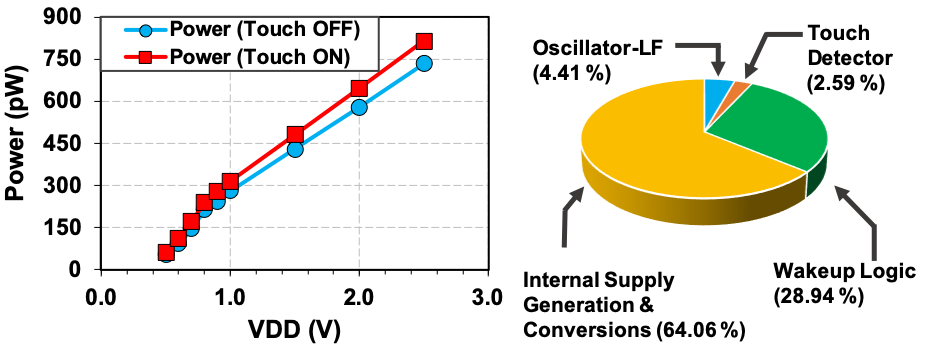}}
  \caption{Characteristics and distribution of components of idle state power.}
  \label{fig:wakeup_characteristics}
\end{figure}

The low-frequency oscillator, the touch detector and the wake-up logic are powered by a 2-transistor version of the voltage reference from \cite{seok_vref_2012} operating from the main supply (VDD). The entire wake-up block consumes sleep power of 735 pW with Osc-LF running at 20.15 Hz at VDD = 2.5 V, as shown in Fig. \ref{fig:wakeup_characteristics}. 

\subsection{Dual-Factor Authentication Unit}

\begin{figure}[!t]
  \centering
  {\includegraphics[width=13 cm]{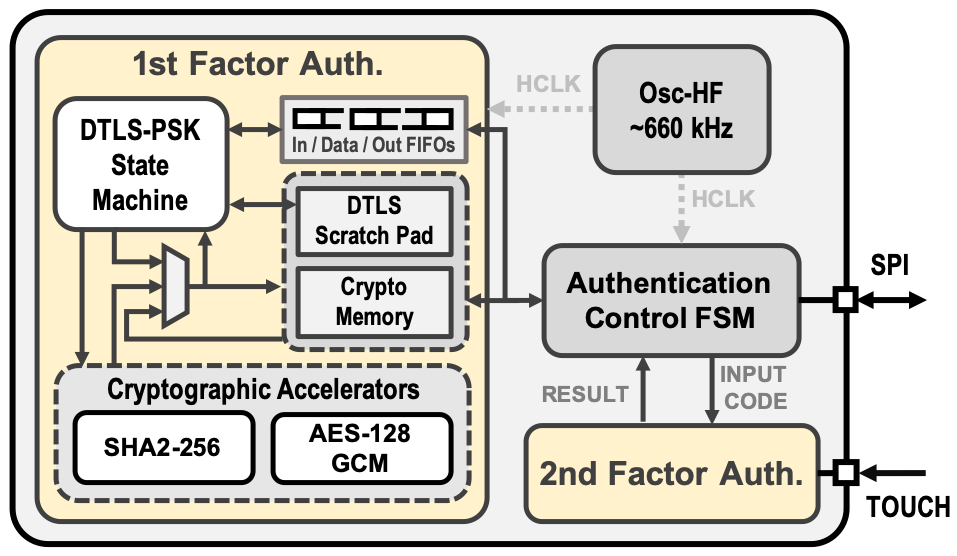}}
  \caption{Architecture of the authentication module consisting of the DTLS-PSK cryptographic accelerator and the second factor verification unit.}
  \label{fig:digital_architecture}
\end{figure}

Fig. \ref{fig:digital_architecture} shows the architecture of the dual-factor authentication block which is clocked by a 660 kHz ring oscillator (Osc-HF) and consists of a cryptographic accelerator, a second factor authentication unit and a control state machine, all powered by the external DC-DC converter (VDC). The cryptographic accelerator implements the DTLS-PSK cipher suite with AES-128-GCM for authenticated encryption/decryption and SHA2-256 for hashing. Embedded software implementations of DTLS typically require 60-80 KB of program memory and 20-30 KB of data memory, which pose prohibitively large area overheads. In order to reduce such overheads of DTLS, \cite{banerjee_isscc_2018} proposed a full hardware implementation of DTLS with elliptic curve cryptography (ECC) cipher suites. Although we do not use any public key cryptography (e.g., ECC), we follow a similar approach for DTLS-PSK by implementing the entire protocol as an optimized micro-coded state machine requiring only 23k-gate logic and 2.75 KB SRAM, thus reducing the area requirement by $\sim 3 \times$. Energy consumption of the DTLS-PSK protocol is further minimized by implementing full data-path AES and SHA round functions in hardware \cite{banerjee_jssc_2019}, with AES-128-GCM and SHA2-256 consuming 14.1 pJ/bit and 5.3 pJ/bit respectively at VDC = 0.87 V. The second factor authentication unit is a simple pattern-matching logic which verifies whether the user’s taps match the pattern sent by the server. The system can also be configured to skip the second factor authentication as a trade-off between security and ease of use.

\begin{figure}[!t]
  \centering
  {\includegraphics[width=13 cm]{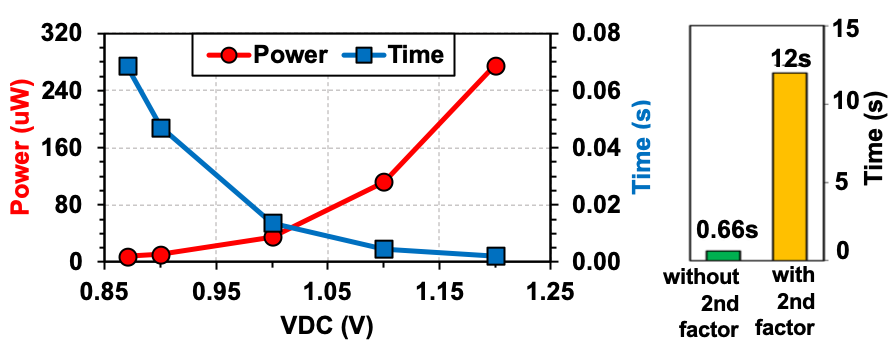}}
  \caption{Measured performance of the authentication unit: (left) scaling of power with supply voltage (right) time of execution of authentication protocol without and with touch-based second factor authentication respectively.}
  \label{fig:measured_architecture}
\end{figure}

The DTLS-PSK-based first factor authentication consumes only 8 $\mu$W at VDC = 0.87 V, which is two orders of magnitude lower than a software implementation. The authentication process takes 12 s and 660 ms respectively in presence and absence of second factor authentication (Fig. \ref{fig:measured_architecture}). Our test chip consumes 5.28 $\mu$J for the first-factor authentication while operating at SPI speed of 125 kbps.

\section{Measurement Results}

\begin{figure}[!b]
  \centering
  {\includegraphics[width=9 cm]{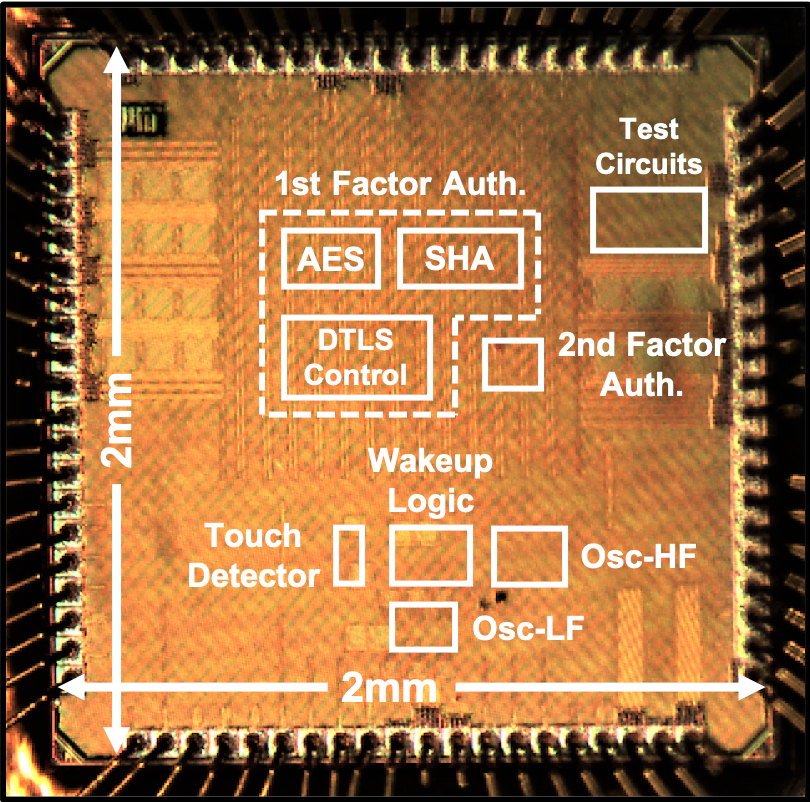}}
  \caption{Chip micrograph (labelled with key design blocks).}
  \label{fig:die_photo}
\end{figure}

\begin{figure}[!t]
  \centering
  {\includegraphics[width=13 cm]{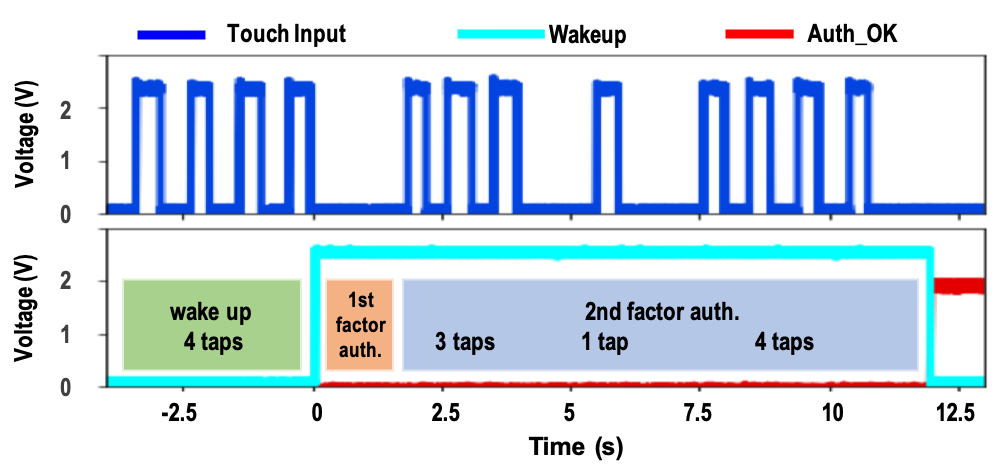}}
  \caption{Waveform of system operation with wake-up, first factor cryptographic authentication and second factor touch-based authentication and final output.}
  \label{fig:system_operation}
\end{figure}

\begin{figure}[!b]
  \centering
  {\includegraphics[width=12 cm]{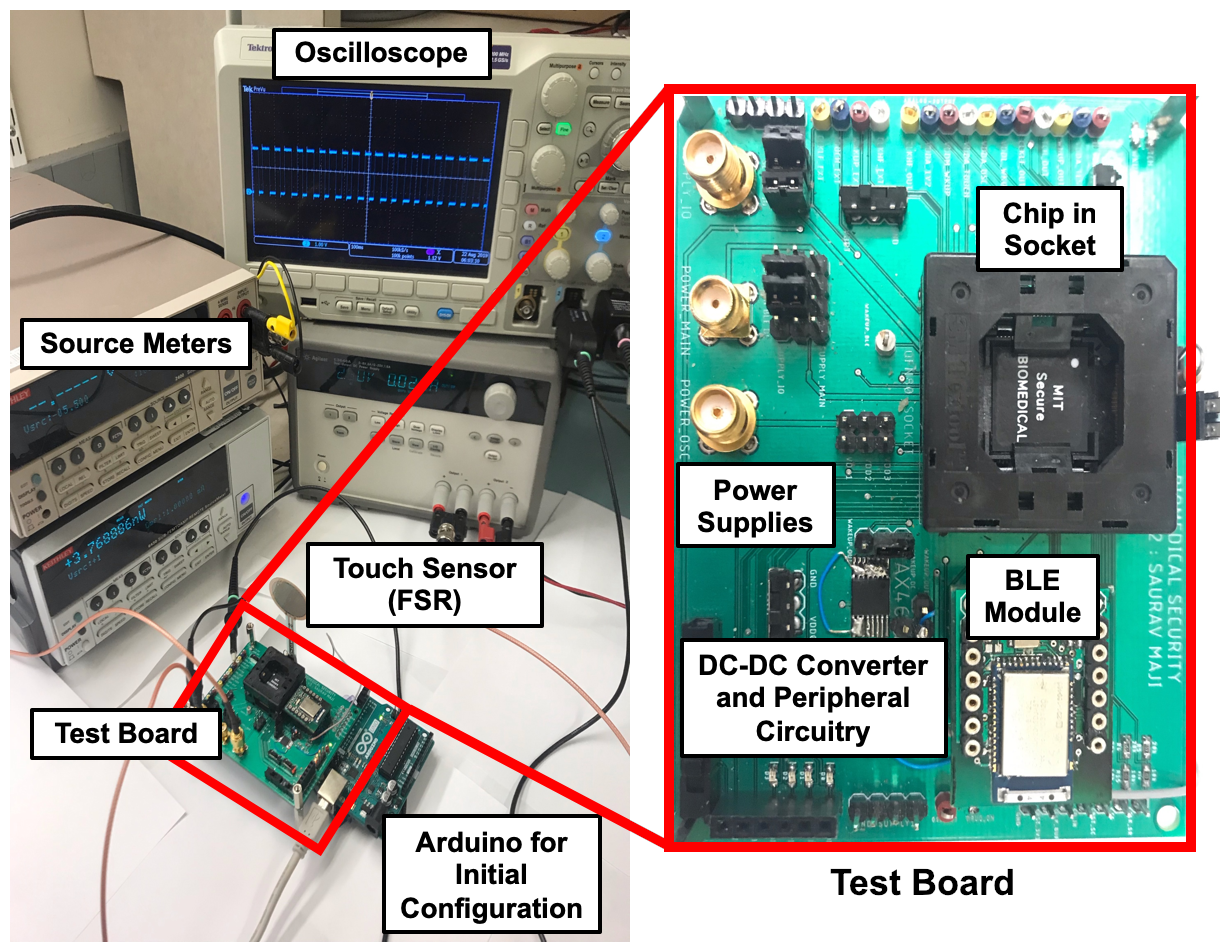}}
  \caption{Measurement setup showing the test chip (in socket), the FSR-based touch sensor, the BLE module, the DC-DC converter and all peripheral components integrated into a test board. An Arduino micro-controller platform is used for initial configuration of the test chip.}
  \label{fig:measurement_setup}
\end{figure}

\begin{figure}[!t]
  \centering
  {\includegraphics[width=8 cm]{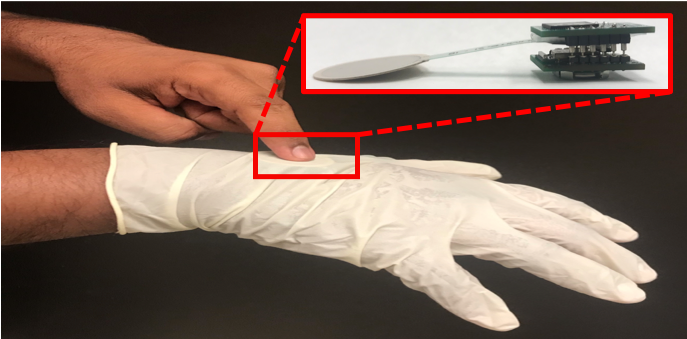}}
  \caption{In-vitro measurement setup (IMD system shown as inset).}
  \label{fig:invitro}
\end{figure}

The test chip, shown in Fig. \ref{fig:die_photo}, was fabricated in TSMC 65nm low-power CMOS process, and all measurements are reported at the following supply voltage combination: VDD = 2.5 V and VDC = 0.87 V, with the low-frequency and high-frequency oscillators operating at 20.15 Hz and 660 kHz respectively. Waveforms of the entire operation with touch input and other relevant signals are shown in Fig. \ref{fig:system_operation}. Power measurements were performed using the setup shown in Fig. \ref{fig:measurement_setup}. In order to demonstrate in-vitro operation of the IMD, we replicated the subcutaneous environment by covering our test system with wet paper and placing it underneath a synthetic glove which acts as a replica of human skin, as shown in Fig. \ref{fig:invitro}. 

\clearpage

\clearpage

\begin{table}[!t]
\renewcommand{\arraystretch}{1.45}
\caption{Comparison with State-of-the-Art Security Processors for Biomedical Applications}
\label{table:comparison}
\centering
\begin{tabular}{|l|C{27mm}|C{27mm}|C{27mm}|C{27mm}|}
\hline
\rowcolor{Gray}
{\textbf{Metric}} & \cite{kim_ecg_2010} & \cite{fan_encrypt_2013} & \cite{hsu_bio_2011} & \textbf{This Work} \\
\hline
Technology (nm) & 180 & 130 & 90 & 65 \\
\hline
Voltage (V) & 0.7 & 1.0 & 0.5 & 0.87 \\
\hline
Operating Frequency & 360 Hz & 847.5 kHz & 25 MHz & 660 kHz \\
\hline
Power & 1.26 $\mu$W & 50.4 $\mu$W & 215 $\mu$W \textsuperscript{a} & 8 $\mu$W \textsuperscript{b} \\
\hline
Area & 2.25 mm\textsuperscript{2} & - & 1.17 mm\textsuperscript{2} & \textbf{315 $\mu$m\textsuperscript{2}} \\
\hline
1st Factor / & Data Encryption & Authentication & Key Ex. / Auth. & DTLS-PSK  \\
Cryptographic Auth. & (AES) & (ECC) & (AES, SHA2, ECC) & (AES, SHA2) \\
\hline
2nd Factor Auth. & No & No & No & \textbf{Yes} \\
\hline
\multicolumn{5}{l}{\textsuperscript{a} Calculated from cryptographic computation energy per cycle and operating frequency in \cite{hsu_bio_2011}} \\
\multicolumn{5}{l}{\textsuperscript{b} Includes both computation and oscillator power}
\end{tabular}
\end{table}

We compare the performance of our authentication unit with previous work on secure biomedical processors in Table \ref{table:comparison}. Our chip is significantly smaller than \cite{kim_ecg_2010,hsu_bio_2011}, and has 1-2 orders of lower power consumption compared to \cite{fan_encrypt_2013,hsu_bio_2011}. Most importantly, ours is the first design to employ dual-factor authentication for security of IMDs.

\section{Conclusion}

In this work, we demonstrate a low-power dual-factor authentication unit to secure implantable medical devices (IMDs). Apart from traditional cryptography using the Datagram Transport Layer Security (DTLS) protocol, our design also employs a touch-based authentication scheme to provide stronger security guarantees. Total power consumption of the dual-factor authentication is only 8 $\mu$W. The touch detector, along with a compact leakage-based oscillator, are used as the always-on wake-up circuitry with 735 pW idle state power consumption. Through circuit-level optimizations, energy-efficient architecture and a novel dual-factor authentication mechanism, this work demonstrates a low-power IC for securing connected biomedical devices of the near future.

\section*{Acknowledgment}

The authors would like to thank Analog Devices Inc. for funding and TSMC University Shuttle Program for chip fabrication support.

\bigskip
\flushleft{\textit{A revised version of this paper was published in the 2020 IEEE Custom Integrated Circuits Conference (CICC) - DOI: \href{https://dx.doi.org/10.1109/CICC48029.2020.9075945}{10.1109/CICC48029.2020.9075945}}}

\bibliographystyle{unsrt}

\begin{thebibliography}{00}

\bibitem{zhang_imd_2014}
M. Zhang, et. al., ``Trustworthiness of Medical Devices and Body Area Networks,'' in \emph{Proc. IEEE}, vol. 102, no. 8, pp. 1174-1188, Aug. 2014.

\bibitem{maji_thesis_2019}
S. Maji, ``Energy-Efficient Protocol and Hardware for Security of Implantable Devices,'' \emph{S.M. Thesis, Massachusetts Institute of Technology}, June 2019.

\bibitem{rescorla_tls1p3_2018}
E. Rescorla, ``The Transport Layer Security (TLS) Protocol Version 1.3,'' \emph{IETF RFC 8446}, Aug. 2018.

\bibitem{eronen_psk_2005}
P. Eronen, et al., ``Pre-Shared Key Ciphersuites for TLS,'' \emph{IETF RFC}, vol. 4279, Dec. 2005.

\bibitem{badra_psk_2009}
M. Badra, ``Pre-Shared Key Cipher Suites for TLS with SHA-256/384 and AES Galois Counter Mode,'' \emph{IETF RFC}, vol. 5487, Mar. 2009.

\bibitem{banerjee_eedtls_2017}
U. Banerjee, et al., ``eeDTLS: Energy-Efficient Datagram Transport Layer Security for the Internet of Things,'' \emph{IEEE GLOBECOM}, pp. 1-6, Dec. 2017.

\bibitem{lin_osc_2009}
Y. Lin, et al., ``A 150pW program-and-hold timer for ultra-low-power sensor platforms,'' in \emph{IEEE ISSCC}, pp. 326-327, Feb. 2009.

\bibitem{nadeau_osc_2016}
P. M. Nadeau, et al., ``Ultra Low-Energy Relaxation Oscillator With 230 fJ/cycle Efficiency,'' in \emph{IEEE J. Solid-State Circuits}, vol. 51, no. 4, pp. 789-799, April. 2016.

\bibitem{wang_osc_2016}
H. Wang, et al., ``A Reference-Free Capacitive-Discharging Oscillator Architecture Consuming 44.4 pW/75.6 nW at 2.8 Hz/6.4 kHz,'' in \emph{IEEE J. Solid-State Circuits}, vol. 51, no. 6, pp. 1423-1435, Jun. 2016.

\bibitem{seok_vref_2012}
M. Seok, et al., ``A Portable 2-Transistor Picowatt Temperature-Compensated Voltage Reference Operating at 0.5 V,'' in \emph{IEEE J. Solid-State Circuits}, vol. 47, no.10, pp. 2534–2545, Oct. 2012.

\bibitem{banerjee_isscc_2018}
U. Banerjee, et al., ``An Energy-Efficient Reconfigurable DTLS Cryptographic Engine for End-to-End Security in IoT Applications,'' in \emph{IEEE ISSCC}, pp. 42-44, Feb. 2018.

\bibitem{banerjee_jssc_2019}
U. Banerjee, et al., ``An Energy-Efficient Reconfigurable DTLS Cryptographic Engine for Securing Internet-of-Things Applications,'' in \emph{IEEE J. Solid-State Circuits}, vol. 54, no. 8, pp. 2339-2352, Aug. 2019.

\bibitem{kim_ecg_2010}
H. Kim, et al., ``A Low Power ECG Signal Processor for Ambulatory Arrhythmia Monitoring System,'' in \emph{IEEE VLSIC}, pp. 19-20, Jun. 2010.

\bibitem{fan_encrypt_2013}
J. Fan, et. al, ``Low-Energy Encryption for Medical Devices: Security Adds an Extra Design Dimension,'' in \emph{ACM/IEEE DAC}, pp. 1-6, Jun. 2013.

\bibitem{hsu_bio_2011}
S. Hsu, et al., ``A Micropower Biomedical Signal Processor for Mobile Healthcare Applications,'' in \emph{IEEE ASSCC}, pp. 301-304, Nov. 2011.

\end{thebibliography}

\end{document}